\documentclass[twocolumn,aps,epsfig,graphicx,showpacs]{revtex4}
\usepackage{graphicx}
\newcommand{\be}{\begin{equation}}
\newcommand{\ee}{\end{equation}}

\newcommand{\bea}{\begin{eqnarray}}
\newcommand{\eea}{\end{eqnarray}}
\newcommand{\bd}{\begin{displaymath}}
\newcommand{\ed}{\end{displaymath}}
\newcommand{\bi}{\begin{itemize}}
\newcommand{\ei}{\end{itemize}}
\newcommand{\bc}{\begin{center}}
\newcommand{\ec}{\end{center}}
\newcommand{\bfl}{\begin{flushleft}}
\newcommand{\efl}{\end{flushleft}}
\newcommand{\bfr}{\begin{flushright}}
\newcommand{\efr}{\end{flushright}}
\newcommand{\f}{\frac}

\def\br{{\bf r}}\def\bR{{\bf R}}
\def\bk{{\bf k}} \def\bK{{\bf K}} \def\bp{{\bf p}}

\def\6{\partial}  \def\b{\beta}
 \def\d{\delta} \def\ve{\varepsilon}

\def\ss{\sigma} 

  \def\D{\Delta}

\def\={\!\!\!&=&\!\!\!}
\def\+{\!\!\!&&\!\!\!+~}
\def\-{\!\!\!&&\!\!\!-~}

\begin{document}

\author{Ionel \c{T}ifrea\cite{adresa} and Michael E. Flatt\'{e}}
\affiliation{Department of Physics and Astronomy and Optical
Science and Technology Center, University of Iowa, Iowa City
52242, USA}
\date{\today}
\title{Electric field tunability of nuclear and electronic spin dynamics due
to the hyperfine interaction in semiconductor nanostructures}

\begin{abstract}
We present formulas for the nuclear and electronic spin relaxation
times due to the hyperfine interaction for nanostructed systems
and show that the times depend on the square of the local density
of electronic states at the nuclear position. A drastic
sensitivity (orders of magnitude) of the electronic and nuclear
spin coherence times to small electric fields is predicted for
both uniformly distributed nuclear spins and for $\d$-doped layers
of specific nuclei. This sensitivity is robust to nuclear spin
diffusion.
\end{abstract}
\maketitle


Traditional semiconductor electronic devices are based on precise
control of the electronic charge distribution using electric
fields, ignoring the spin degrees of freedom of the electrons.
Similar control over an electron's spin may lead to the
development of new electronic devices with improved performance
and new functionality.\cite{book,wolf} Electronic spin coherence
times exceed 100 ns at low temperatures in GaAs,\cite{om1} and
nuclear spin coherence times can exceed 1 s in GaAs quantum wells
(QW).\cite{berg,NMR1,smet} Because of these long coherence times,
nuclear spins are also candidates for spin-based
devices.\cite{kane}

A natural way to control both electronic and nuclear spins would
rely on magnetic fields. However, high magnetic fields are
difficult both to achieve and to change rapidly. Furthermore,
detection of nuclear magnetic resonance (NMR) signals from samples
of reduced dimensionality is limited by the low nuclear
polarization achievable with standard techniques.\cite{NMR1}
Studies in semiconductor quantum
wells,\cite{NMR1,NMR2,harley,salis1} show that a strong local
magnetic field and high nuclear polarization emerge as a
consequence of optically-induced dynamical nuclear polarization
(DNP)\cite{over1}  via the hyperfine interaction. Kawakami {\it et
al.},\cite{kawa1} have further demonstrated ``imprinting" of
nuclear spin polarization from adjacent ferromagnetic layers. Smet
{\it et al.},\cite{smet} have manipulated nuclear spins by
electrically tuning the electron density in a QW across  a Quantum
Hall ferromagnet transition; the electric field tunes the nuclear
spin relaxation time by changing the spectrum of collective mode
excitations. Polarization of nuclei has also been predicted to
alter electronic decoherence dynamics in quantum dots.\cite{loss}
Hence, the electronic-nuclear spin interaction is of major
interest, with implications for both electronic and nuclear spin
lifetimes.\cite{over2}

Here we derive general formulas applicable to nanostructures for
the nuclear and electronic spin relaxation and decoherence times,
$T_1$ and $T_2$, from the hyperfine interaction. The central
physical quantity is the electronic local density of states
(ELDOS) at the nuclei. We re-analyze the measurements of
Ref.~\cite{NMR1} using these formulas to obtain new values of the
hyperfine coupling in GaAs QW's. We predict that the dominant
process for nuclear $T_1$ in these QW's (and $T_2$ in others) can
be tuned with an electric field by modifying the ELDOS at
particular locations. For a parabolic QW electric-field tuning of
nuclear spin relaxation by many orders of magnitude is possible,
at temperatures considerably higher than in Ref. \cite{smet} and
despite nuclear spin diffusion. The calculations of nuclear spin
diffusion properly consider the ELDOS and inhomogeneous nuclear
magnetization and indicate non-exponential long-time nuclear
dynamics.

We assume nuclei are polarized through DNP and most of our
calculations are performed at 30K, where DNP is very efficient
with typical laboratory magnetic fields (although tunability of
$T_1$ and $T_2$, in principle, extends to much higher
temperatures). In GaAs QW's the nuclear $T_1$ is dominated by the
hyperfine interaction; however nuclear dipolar interactions limit
$T_2$ to  $10^{-4}$~s. The electronic $T_1$ and $T_2$ in GaAs QW's
are dominated by other processes.  Therefore our specific
predictions focus on control of the nuclear $T_1$. The general
equations, however, are valid for describing the tuning of nuclear
$T_2$ and electronic $T_1$ and $T_2$ in situations where the
hyperfine interaction dominates those times. At the end of this
Letter we propose several such situations.

For GaAs QW's we propose two different experimental configurations
to demonstrate the electric field tunability of the nuclear $T_1$.
The same approaches can be used to tune nuclear $T_2$ and
electronic spin decoherence in other material systems. In the
first configuration, the $T_1$ of Ga and As nuclei in the
nanostructure depends on the occupancy of conduction subbands,
decreasing stepwise as the number of occupied conduction subbands
(and hence the density of states) increases. Manipulation of the
QW density, and implicitly the number of occupied subbands, can be
accomplished with a gate voltage, permitting the manipulation of
$T_1$.  In the second configuration, a single $\d$-doped layer of
a different material (such as In) is inserted at a specific
position. The tunability of $T_1$ of these nuclei comes from the
change in the electronic wave functions due to the applied
electric field.


Our analysis of the electronic and nuclear spin relaxation times
due to the hyperfine interaction in low dimensional systems
follows in spirit the calculation by Overhauser\cite{over2} for
bulk metals, but now includes new effects due to the
nanostructure. The interaction Hamiltonian can be written as
\be\label{intham}
H=\f{8\pi}{3}\;\b_e\b_n\left({\vec \ss_n }\cdot{\vec
\ss_e}\right)\;\d(\br-\br_n)\;,
\ee
where $\b_n$ and $\b_e$ are the nuclear and electron magnetic
moments,  and ${\vec \ss_n}$ and ${\vec \ss_e}$ are the Pauli spin
operators for the nucleus and electron. The argument of the delta
function, $\br-\br_n$, represents the relative distance between
the nuclear and electronic spins. The main effect of this
Hamiltonian is a spin-flip process involving both the electronic
and nuclear spins, which we evaluate using Fermi's golden rule.

The time dependence of the electronic magnetization is
\be\label{Dfinal}
\f{dD}{dt}=\f{D_0-D}{T_{1e}}+G\f{\D_0-\D}{T_{1n}}\;,
\ee
where $D$ and $\D$ are the electronic and nuclear
magnetization with $D_0$ and $\D_0$ their equilibrium values, and
$G=2I(I+1)(2I+1)/3$ ($I$ represents the nuclear spin magnetic
number). The electronic ($T_{1e}$) and nuclear ($T_{1n}$)
relaxation times for general nanostructures and weak spin
polarization are thus
\be\label{te}
T^{-1}_{1e}=\f{1}{V}\sum_{\br_n}
\f{1024\pi^3\;\b_e^2\;\b_n^2\;\int d\ve \; A_e^2(\br_n,\ve)\;
f'_{FD}(\ve)}{9\hbar\;(2I+1)\;\int d\br\;d\ve\; A_e(\br,\ve)\;
f'_{FD}(\ve)}
\ee
and
\be\label{tn}
T^{-1}_{1n}(\br_n)=\f{512\pi^3\;\b_e^2\;\b_n^2\;k_BT\;\int d\ve\;
A_e^2(\br_n,\ve)\; f'_{FD}(\ve)} {3\hbar\;I(I+1)(2I+1)}\;,
\ee
where
\be\label{eldos}
A_e(\br_n,\ve)=\sum_{m} |\psi_{m}(\br_n)|^2\;\delta(\ve - E_{m}).
\ee
Here $A_e(\br_n,\ve)$ is the ELDOS ($m$ labels the state, and
$\psi_{m}(\br_n)$ the electron wave function of that state at the
nucleus), $f_{FD}(\ve)$ the Fermi-Dirac distribution function, and
$T$ the temperature. If there is no energy bottleneck for the
electron (e.g. there is none in QW's), the transverse spin
decoherence rate $T_2^{-1}$ from this mechanism is equal to
$T_1^{-1}$. According to Eqs. (\ref{te})-(\ref{eldos}), the
electronic and nuclear spin relaxation times will depend on the
position of the nuclei. $T_{1e}$ is temperature independent,
suggesting that it is possible for the hyperfine interaction to
dominate $T_{1e}$ at low temperatures, for the relaxation times
corresponding to other electronic mechanisms increase as the
temperature decreases.\cite{over2,opticalbook}

For a QW the system's dispersion relations are
quasi-two-dimensional; therefore, the electronic wave functions
can be written as a product between an envelope function,
$\phi(z)$, and a Bloch function, $u(\br)$, such that
$\psi_{j\bK}(\br_n)=\exp{[i\bK\cdot\bR]\;\phi_j(z)\;u(\br_n)}$.
For this situation $A_e(\br_n,\ve)=\sum_{j} |\phi_{j}({\bf
z}_n)|^2\;N_{2D}\Theta(\ve - E_{j(\bK=0)})$, where $N_{2D}$ is the
density of states for a two-dimensional electron gas and $\Theta$
is the Heaviside step function. Available experimental data for
the nuclear spin relaxation time of a GaAs/Al$_{0.1}$Ga$_{0.9}$As
QW \cite{NMR1} allow us to extract the value of the conduction
band Bloch function, $|u(\br_n)|^2 = 5.2\times 10^{25}$~cm$^{-3}$.
This value compares well with $|u(\br_n)|^2 = 5.8\times
10^{25}$~cm$^{-3}$ extracted from bulk GaAs in
Ref.~\cite{paget1977}.  $\phi_j(z)$ is evaluated using a
fourteen-band $\bk\cdot\bp$ calculation.\cite{thesis} We consider
now two different systems: a square GaAs QW ($L=75$\AA) confined
within two barriers of Al$_{0.4}$Ga$_{0.6}$As and a parabolic
Al$_x$Ga$_{1-x}$As QW ($L=1000$\AA) confined within two barriers
of Al$_{0.4}$Ga$_{0.6}$As. The parabolic QW is obtained by
gradually varying the Al concentration, $x$, of Al$_x$Ga$_{1-x}$As
layers from 0.4 in the two barriers to 0.07 in the center of the
QW.

\begin{figure}[b]
\includegraphics[width=7cm]{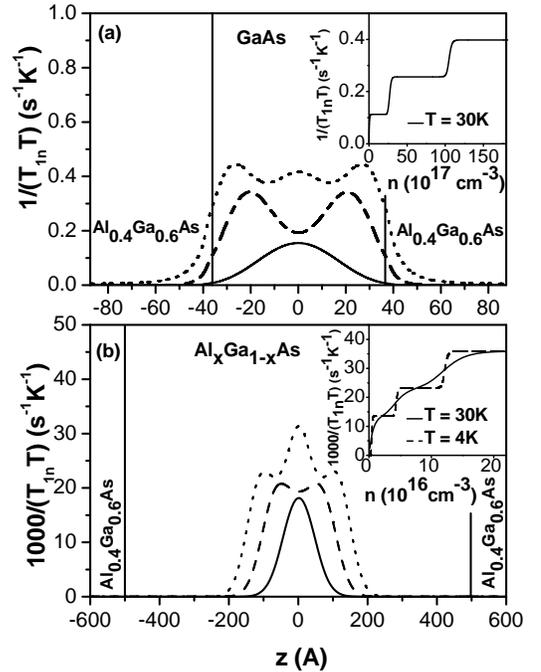}
%
%
\caption{The nuclear spin relaxation rate as function of the
position in the QW for different conduction subband occupancy at
$T=30K$ (full line-single subband occupancy, dashed line-double
subband occupancy, and dotted line-triple subband occupancy).
Inset: initial nuclear spin relaxation rate for different subband
occupancy. ((a) Square GaAs QW. (b) Parabolic Al$_x$Ga$_{1-x}$As QW).}%
\label{fig1}
\end{figure}

In Fig. \ref{fig1} we present the position dependence of the
relaxation times for the square GaAs (Fig. \ref{fig1}a) and
parabolic Al$_x$Ga$_{1-x}$As (Fig. \ref{fig1}b) QW's for different
conduction band occupancy. The shape of the curves describing the
$T_{1n}(z)$ are similar for the two considered situations. An
initial nuclear polarization obtained by DNP will be
inhomogeneous, and for short times will be proportional to
$T^{-1}_{1n}(z)$, so for one occupied subband the initial nuclear
magnetization $m(z,t=0)\propto |\phi(z)|^4$. The {\em initial}
$T_{1n}$ for the {\it total} Ga and As nuclear magnetization
initialized this way is plotted in the insets of Fig. \ref{fig1}
as a function of electron density. Note that as the electron
density in the QW increases, the number of occupied conduction
subbands will increase, and as a consequence the $T_{1n}$  will
decrease stepwise even for these uniformly distributed Ga and As
nuclei. For the parabolic QW, where the energy difference between
the minimum of two consecutive conduction subbands is about 15
meV, thermal smearing of the Fermi function at T=30K (solid line)
will suppress the stepwise shape of the initial $T_{1n}$. However,
at T=4K, where the Fermi function is sharper (dashed line), the
stepwise dependence of the $T_{1n}$ is observable. Application of
an electric field across a QW can also tilt the confining
potential. The direct dependence of both electron and nuclear spin
coherence and relaxation times on the electronic envelope function
suggests that control of spin relaxation times can thus be
achieved by using such an external electric field ${\bf E}$. Salis
{\it et al.} \cite{salis2} suggested that the wavefunction shift
with ${\bf E}$, and hence the electrical control of spin
coherence, is particularly effective in a shallow parabolic QW.

We now consider the effects of shifting the electronic envelope
wave functions to overlap different parts of the initial polarized
nuclear population (different positions), {\it and nuclear spin
diffusion}, by tracking the polarizations at the different
positions as a function of time and then summing them to track the
time dependence of the total nuclear polarization. For this and
all subsequent calculations  we consider electron densities where
only the first subband is occupied. $m(z,t)$ can be obtained by
solving
\be\label{average}
\f{d m(z,t)}{d t}=D\f{\6^2m(z,t)}{\6z^2}-\f{m(z,t)}{T_{1n}(z)}\;,
\ee
where $D$ represents the diffusion constant, whose value is of the
order of 10$^3$ A$^2$/s for GaAs systems.\cite{paget} Our results
indicate that the longer-time dynamics of the magnetization will
be non-exponential.

\begin{figure}[t]
\centering
\scalebox{0.4}[0.45]{\includegraphics*{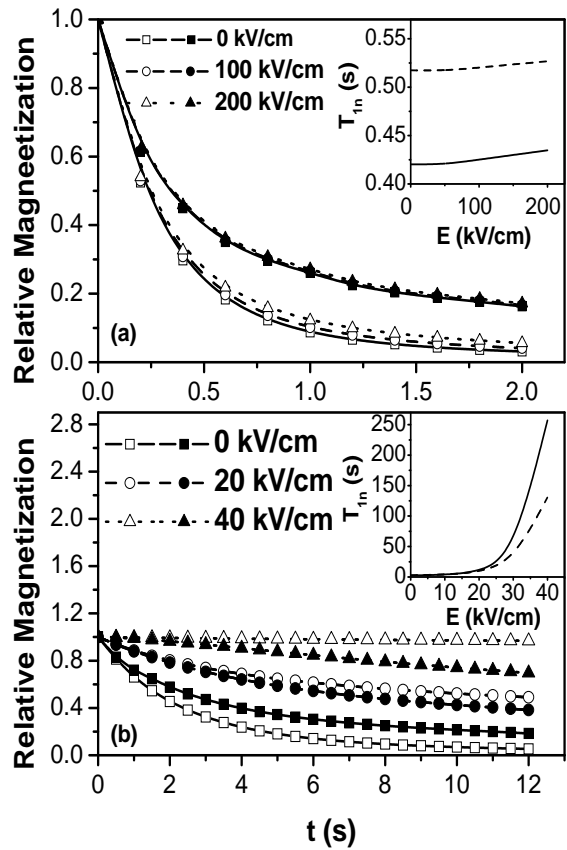}}
%
%
\caption{The total relative nuclear magnetization as function of
time for different values of the applied electric field at $T=30
K$ in the presence (full symbols) and absence (open symbols) of
diffusion. Inset: total nuclear spin relaxation time as function
of the electric field in the presence (full line) and the absence
(dashed line) of diffusion ((a) Square GaAs QW; (b) Parabolic Al$_x$Ga$_{1-x}$As QW).}%
\label{fig2}
\end{figure}

\begin{figure}[t]
\centering
\scalebox{0.4}[0.45]{\includegraphics*{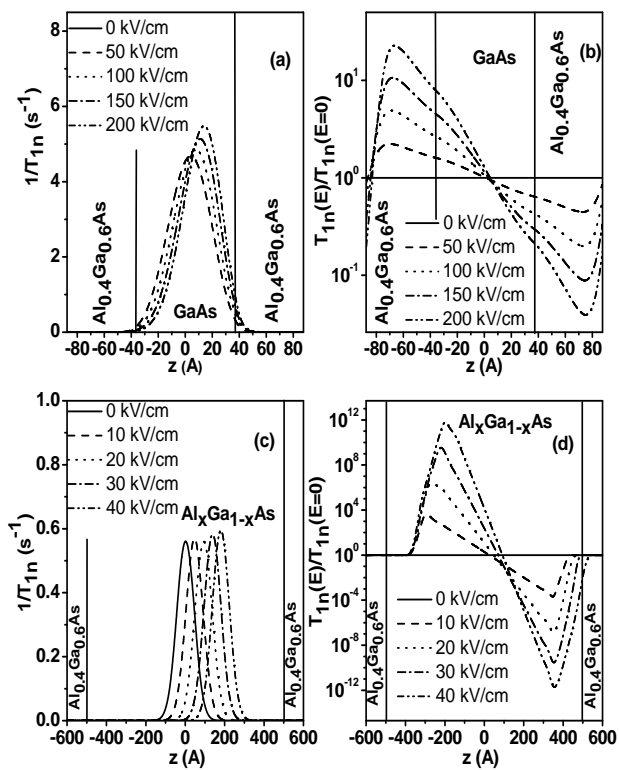}}
%
%
\caption{The nuclear spin relaxation rate and the ratio of the
relaxation times in the presence and the absence of the electric
field as function of the position in the QW for different values
of the applied electric field at $T=30K$. ((a) and (b) Square GaAs
QW.
(c) and (d) Parabolic Al$_x$Ga$_{1-x}$As QW).}%
\label{fig3}
\end{figure}

In Fig. \ref{fig2} we plot, for Ga and As nuclei which have been
polarized via DNP at ${\bf E}=0$, the time dependence of the total
QW's nuclear magnetization for different values of the applied
electric fields in the presence and the absence of spin diffusion.
The inset shows the field dependence of the total initial nuclear
spin relaxation time extracted as the first derivative of the
magnetization at $t=0$~s. The diminished overlap of the electron
envelope function with the region of polarized nuclei reduces the
relaxation rates [shown in Fig. \ref{fig3}(a) and (c)].
Magnetization decay in the square QW (Fig \ref{fig2}a) is almost
unaffected by the electric field, whereas for the parabolic QW
(Fig. \ref{fig2}b) a large increase of the relaxation time is
obtained even in small electric fields. In the presence of nuclear
spin diffusion the effect of the electric field is reduced; in the
parabolic QW, however, one can still see a significant difference
between relaxation times at different applied electric fields.
Recent measurements of nuclear spin diffusion in AlGaAs barriers
indicated diffusion constants an order of magnitude smaller than
in GaAs.\cite{harley2} This suggests the tunability in the
parabolic QW may be even more robust to diffusion than shown in
Fig. {\ref{fig2}}.

An even more precise level of electric field control is possible
in structures which have been intentionally $\d$-doped with a
layer of different nuclei, such as In. For such a structure
$T_{1n}^{-1}$ depends on the position of the $\d$-doped layer
according to Fig. \ref{fig3}a, assuming the Bloch function on In
is the same as that on Ga. Although $T_{1n}$ for this layer could
vary considerably, in a GaAs host $T_{2n}$ would not because of
transverse spin diffusion to or from the host nuclei. In Fig.
\ref{fig3}b we plot the ratio of the spin relaxation times in the
presence and absence of an applied electric field as a function of
the position along the growth direction for the square QW. We can
see that the effect of the electric field is strongest within the
two barriers. The effect of the electric field is far greater for
the parabolic QW; the spin relaxation times increase four orders
of magnitude for an electric field as low as 10kV/cm (Fig.
\ref{fig3}d), and this increase occurs in regions of large initial
nuclear polarization. If the $T_{1n}$ from the hyperfine
interaction is made sufficiently long, eventually the total
$T_{1n}$ will come to be dominated by the $\sim 10$~min $T_{1n}$
time\cite{spinphonon} from spin-phonon interactions.

From Eqs. (\ref{te})-(\ref{eldos}) and $T_{1n}$ from Ref.
\cite{NMR1} we estimate the spin relaxation time via the hyperfine
interaction for the {\em electron} in both the square and
parabolic QW's of Fig. \ref{fig1}. For both structures we obtain
$T_{1e}\approx 10^{-5}$s for a single occupied subband. For
uniformly distributed nuclei the electric field dependence of
$T_{1e}$ is small, except for the density dependence (which is the
same as shown in the Fig. \ref{fig1} inset for $T_{1n}$). The
electronic relaxation time is $T_{1e}\sim 10 ^{-7}$s from other
processes; however if these other processes could be suppressed
times of $10^{-5}$s might be observable. The influence of a
$\d$-doped layer of nuclei on $T_{1e}$ could also be electric
field tuned, with the same behavior as $T_{1n}$ in Fig.
\ref{fig3}.

We conclude by describing how to reduce competing processes for
both the nuclear and electron spin coherence times. We have
considered In as the $\d$-doped layer of nuclei in the GaAs QW.
Although the different resonant frequency will limit the effect on
the In $T_{1n}$ of spin diffusion to the Ga and As nuclei, the host
nuclei could significantly reduce the $T_{2n}$ through dipole-dipole
coupling. Another choice of QW, ZnCdSe/ZnSe, can be grown entirely
from spin-0 nuclei, hence a $\d$-doped Mn layer in this structure
should have a $T_{2n}$ dominated by the tunable hyperfine
interaction.

Electrons in either GaAs or ZnCdSe QW's may have $T_{1e}$'s
limited by spin-orbit interaction. Si QW's in SiC (or
SiO$_2$)\cite{carrier}, however, may have both spin-0 nuclei and
weak spin-orbit interaction. Thin Si layers in these QW structure
can have a direct gap, so these layers could be probed or pumped
optically. The electron spin coherence times could then be
dominated by interactions with the $\d$-doped nuclei. In these
QW's a good choice for a $\d$-doped nucleus the spin-1/2 Si
nucleus.

We would like to acknowledge D. D. Awschalom, W. H. Lau, D. Loss,
A.V. Khaetskii, J. M. Kikkawa and J. M. Tang for helpful
discussions. Our work was supported by DARPA/ARO DAAD19-01-1-0490.

\end{document}